\documentclass[12pt]{article}

\usepackage{epsfig}
\usepackage{graphicx}

\newcommand{\be}{\begin{equation}}
\newcommand{\ee}{\end{equation}}
\newcommand{\bea}{\begin{eqnarray}}
\newcommand{\eea}{\end{eqnarray}}
\newcommand{\ba}{\begin{array}}
\newcommand{\ea}{\end{array}}
\newcommand{\nn}{\nonumber
\\}

\font\mybb=msbm10 at 10pt
\def\bb#1{\hbox{\mybb#1}}

\def\bZ {\bb{Z}}
\def\bR{\bb{R}}

\begin{document}


\begin{titlepage}
\rightline{UB-ECM-PF-04/23}
\rightline{DAMTP-2004-89}
\rightline{hep-th/0408220}
\vfill

\begin{center}
\baselineskip=16pt
{\Large\bf Cosmology as Relativistic Particle Mechanics: \\
From Big Crunch to Big Bang}
\vskip 0.3cm
{\large {\sl }}
\vskip 10.mm
{\bf ~J.G. Russo$^{*}$ and P.K. Townsend$^{*,\dagger}$ 
}
\vskip 1cm
{\small
$^*$
Instituci\'o Catalana de Recerca i Estudis
Avan\c cats,
\\
Departament ECM, Facultat de F\'{\i}sica,
\\
Universitat de Barcelona,
Diagonal 647,\\
E-08028 Barcelona,
Spain.\\
}
\vspace{6pt}
{\small
$^\dagger$
Department of Applied
Mathematics and Theoretical Physics \\
Centre for Mathematical
Sciences, University of Cambridge\\
Wilberforce Road, Cambridge, CB3
0WA,
UK.\\
}

\end{center}
\vfill

\par
\begin{center}
{\bf
ABSTRACT}
\end{center}
\begin{quote}

Cosmology can be viewed as geodesic motion in an appropriate metric on an `augmented' target space; here we obtain these geodesics  from  an effective relativistic particle action.  As an application, we find some exact (flat and curved) cosmologies for models  with $N$ scalar fields taking values in a hyperbolic target space for which the augmented target space is a Milne universe. The singularities of these cosmologies correspond to points at which the particle trajectory crosses the Milne horizon, suggesting a novel resolution of them, which we explore via the Wheeler-DeWitt equation.
\end{quote}
\end{titlepage}


\setcounter{equation}{0}

\section{Introduction}

The general (low-energy) Lagrangian for  $d$-dimensional gravity coupled to $N$ scalar fields 
$\phi^\alpha$ taking values in a target space with (positive definite) metric $G_{\alpha\beta}$ is 
\be\label{gravscalar}
L= \sqrt{-g} \left( {1\over4}R -{1\over2} G_{\alpha\beta} (\phi)\, \partial\phi^\alpha \cdot \partial\phi^\beta - V(\phi)\right),
\ee
where  $V$ is a potential energy  function on the target space.  Here we discuss cosmological solutions of the equations of motion of this model. Assuming homogeneity and isotropy, spacetime coordinates can be found such that the spacetime metric takes the FLRW form
\be\label{FLRW}
ds^2 = -dt^2 + S^2(t)\, d\Sigma_k^2,
\ee
where $S(t)$ is the (positive) scale factor and $\Sigma_k$ represents the (d-1)-dimensional spatial sections of constant curvature $k$, which we normalize so that $k=0,\pm1$.  Note that
\be\label{sqrtg}
\sqrt{-g} = \eta \, {\rm vol}_k \qquad \eta := S^{d-1}, 
\ee
where ${\rm vol}_k$ is the normalized volume element of a spatial section. Homogeneity and isotropy also imply that the scalar fields are functions only of time. Thus, a cosmological solution is a trajectory in an `augmented target space' (alias mini-superspace) parametrized by $N+1$ coordinates, which can be chosen to be $\Phi^\mu=(\eta,\phi^\alpha)$. For $V=0$ and $k=0$ it has long been appreciated  \cite{Dewitt:1967ub, Gibbons:1986xp} that the trajectories in this space are null geodesics  with respect to the Lorentzian signature metric
\be\label{augmetric}
G_{\mu\nu} d\Phi^\mu d\Phi^\nu =  - \alpha_h^{-2} \left({d\eta\over \eta}\right)^2 + 
 G_{\alpha\beta}\, d\phi^\alpha d\phi^\beta,  
\ee
where $\alpha_h$ is the `hypercritical' coupling constant, the general significance of which was explained in \cite{Townsend:2004zp} and which in our conventions takes the numerical value 
\be
\alpha_h = \sqrt{{2\left(d-1\right)\over d-2} }. 
\ee
It was shown in \cite{Townsend:2004zp} that $k=0$ trajectories for non-zero $V$ are {\it also} geodesics in the augmented target space, but with respect to the conformally rescaled metric 
\be
\tilde G_{\mu\nu} = \Omega^2 G_{\mu\nu}, \qquad
2\Omega^2 = \cases{  \eta^2 \, |V| & $V\ne0$ \cr  \eta^2 & V=0.}
\ee
The geodesic is timelike if $V>0$, spacelike if $V<0$ and null if $V=0$. This observation is particularly useful if the target space is flat, because $G_{\mu\nu}$ is then the Minkowski metric and $\tilde G_{\mu\nu}$ is conformally equivalent to Minkowski space. This was  the case analysed in detail in \cite{Townsend:2004zp}. 

Cosmologies with $k\ne0$ were interpreted in  \cite{Townsend:2004zp} as projections of geodesics from a `doubly-augmented' target space of dimension $N+2$. For $k=-1$ this reformulation is particularly attractive because, in this case, the `doubly-augmented' target space has Lorentzian signature, 
and the geodesics are again either timelike, spacelike or null, according to the sign of $V$.
One result of this paper is a geometrical interpretation of the projection involved in this construction as Scherk-Schwarz dimensional reduction;  formally, the same applies to $k=1$ cosmologies but the extra dimension in this case is timelike. An alternative formulation of  $k=\pm1$ cosmologies as geodesics, which we explain here, attributes the effect of the spatial curvature to an additional  term in an {\it effective} potential, to which the $k=0$ methods may then be applied.  

In effect, the results just summarized reduce the study of FLRW cosmologies to  free particle mechanics!  Actually, this should not come as such a surprise because the Maupertuis-Jacobi principle of classical mechanics makes a similar claim for many dynamical systems. Consider a particle of unit mass moving on a Riemannian space with metric $g_{ij}$ in coordinates $Q^i$, in a potential $V(Q)$. Its Lagrangian may be written in time-reparametrization invariant form as 
\be
L =  P_i \dot Q^i - H\dot T  + e\left(H - {1\over 2}g^{ij}P_iP_j - V\right)
\ee
where $P_i$ is the particle's momentum, the Lagrange multiplier $e$ is the `gauge field' for the time-reparametrization invariance, and the overdot indicates differentiation with respect to an arbitrary parameter. Note that the variable $H$ can also be interpreted as a Lagrange multiplier for the constraint $e=\dot T$, so the standard gauge choice $\dot T=1$ (for which $H$ becomes the Hamiltonian) is equivalent to the gauge choice $e=1$. However, we may make the alternative gauge choice $\dot T = \left(H-V\right)^{-1}$, which is equivalent to 
\be
e^{-1}= H-V. 
\ee
Elimination of the momentum then yields the Lagrangian
\be
L=  T\dot H + {1\over2}\left(H-V\right) g_{ij}\dot Q^i\dot Q^j, 
\ee
where we have discarded a total time derivative. Now $T$ is a Lagrange multiplier, and the constraint it imposes can be solved by setting $H=E$ for some constant (energy) $E$. We then arrive at the Lagrangian
\be
L= {1\over2}\tilde g_{ij}\dot Q^i\dot Q^j, \qquad \tilde g_{ij} := \left(E-V\right) g_{ij}\,  .
\ee
The equations of motion of this Lagrangian are the equations for
affinely-parametrized geodesics in the conformally rescaled metric
$\tilde g$, and the affine parameter $\tau$ is related to the time
parameter $t$ corresponding to the standard gauge choice by the
equation $d\tau= (E-V)dt$. Note that the conformal factor goes to zero at the boundary of any classically forbidden region for which $E<V$, although a quantum mechanical particle may tunnel through these regions. 

Cosmology as geodesic motion can be deduced in an analogous manner from an 
 effective {\it relativistic} point particle Lagrangian, in which the Friedmann constraint arises as a mass-shell constraint. This idea, and its connection to the Maupertuis-Jacobi principle has a long 
 history\footnote{We thank Lee Smolin for correspondence on this point, and for bringing the Maupertuis-Jacobi princple to our attention.}. The  principle first arose in General Relativity in connection with the `problem of time'  \cite{BSW} ; see also \cite{Brown:1989ne, Barbour:2000qg}, for example, and  \cite{Greensite:1995ep,Carlini:1996fs} for a more recent discussion that extends the principle 
to include scalar fields. The application to cosmology appears to have emerged in several steps, beginning with the work of DeWitt cited above. 

The main purpose of this paper is to apply these ideas to some
specific models in which the $N$ scalar fields take values in a
hyperbolic target space. It turns out that for a particular radius of
this target space the metric on the augmented target space with
respect to which the cosmological trajectories are geodesics is the
metric of the Milne universe. As this is just a region of Minkowski
space, the geodesics are just straight lines. We present two different
realizations of this idea. One, for which the exact solution for all
fields can be found in terms of FLRW time, involves $k=0$ cosmologies in a model
with constant potential; this case generalizes a model of
$(2+1)$-dimensional gravity \cite{Waldron:2004gg} (see also 
\cite{Pioline:2002qz}) that came to our attention while we were 
writing up our work. The other case (which requires a different radius
for the hyperbolic target space) occurs for $k\ne0$ cosmologies 
with $V=0$.  

The ubiquity of the Milne metric for gravity/scalar models with
hyperbolic target spaces is remarkable (and goes beyond the particular
cases discussed here). Perhaps the most important aspect of this is
the insight it gives into the nature of cosmological singularities (of
these models, at least). Indeed, it suggests how a big crunch
singularity might be resolved, leading to a subsequent re-expanding
phase, as has often been suggested in the past. We explore this possibility in the context of the Wheeler-DeWitt equation of quantum cosmology, which is just the wave-equation of the associated relativistic particle.

\section{Flat cosmologies from particle mechanics}

Consider the time-reparametrization invariant particle 
Lagrangian
\be\label{effL}
L= p_\mu \dot \Phi^\mu - 
{1\over2}e\left(\eta^{-2}G^{\mu\nu} p_\mu p_\nu + 2\, V\right), 
\ee
where $e$ is the `einbein', or lapse function, that also acts
as a Lagrange multiplier for a `mass-shell'  constraint.  
The
overdot indicates differentiation with respect to an arbitrary time
parameter, which can be related to the FLRW time $t$ only after a
choice of gauge has been made. Comparison of the potential term in
this effective particle Lagrangian with its appearance in the original
Lagrangian (\ref{gravscalar}) shows that $e$ is essentially the spacetime volume
density $\sqrt{-g}$. From (\ref{sqrtg}) we see, for $k=0$ universes in
standard FLRW coordinates, that $\sqrt{-g}=\eta$. It follows that the
choice of FLRW time $t$ corresponds to the gauge choice $e=\eta$ in
the effective particle Lagrangian. However, let us now consider  the
alternative gauge choice
\be\label{gaugechoice}
e^{-1} = 2\eta^{-2}
\Omega^2 = \cases{ |V| & $V\ne0$ \cr 1 & $V=0$}.
\ee
This gauge
choice corresponds to a choice of time variable $\tau $ that is
related to the FLRW time $t$ by\footnote{What we call $\tau$ here was
  called $\hat t$ in \cite{Townsend:2004zp}.}
\be\label{tthat}
d\tau
= 2\eta^{-1}\Omega^2\, dt. 
\ee
For non-constant $V$ this defines
$\tau $ very 
implicitly because $V$ depends on $t$ through its dependence 
on the scalar fields, and {\it their} time-dependence is itself 
dependent on the cosmological trajectory. However, the equations 
of motion are very simple in this gauge. The momentum variables can 
be solved to yield
\be
p_\mu = 2 \tilde G_{\mu\nu}\dot\Phi^\nu, 
\ee
and the Lagrangian then becomes
\be
L= \tilde G_{\mu\nu} \dot \Phi^\mu \dot \Phi^\nu. 
\ee
Solutions of the Euler-Lagrange equations are affinely-parametrized geodesics in the metric $\tilde G$, while the mass-shell constraint is equivalent to\footnote{Although it might appear that there is some discontinuity on geodesic  trajectories for which the potential
changes sign, this is an artifact of the gauge choice (\ref{gaugechoice}), which was made to ensure an affine parametrization of the geodesic. The form of the initial Lagrangian (\ref{effL}) shows that there is no physical discontinuity.}
\be
\tilde G_{\mu\nu}\dot\Phi^\mu\dot\Phi^\nu  = \cases{-{\rm sign}\, V & $V\ne0$ \cr 0 &$V=0$.}
\ee
These are precisely the equations shown in  \cite{Townsend:2004zp} to correspond to $k=0$ cosmologies in an affine time variable $\tau $ that is related to the FLRW time $t$  by (\ref{tthat}). 

Note that  $\Omega^2\rightarrow 0$  as $\eta\rightarrow 0$. In our earlier discussion of the Maupertuis-Jacobi principle we noted that the conformal factor vanishes at the boundaries of regions in space that are classically forbidden, but that a particle could pass through these regions by quantum mechanical tunnelling. This suggests that big-bang or big-crunch singularities might be similarly viewed as 
boundaries of classically forbidden regions that could be traversed via tunnelling. We will return to this idea after we have introduced a class of models in which its application is most likely to prove 
feasible. 

\section{Hyperbolic sigma model cosmology}
\label{sect.hyper}
We now consider the special case of a hyperbolic target space. 
Specifically, we take the target space metric to be 
\be
G_{\alpha\beta} \, d\phi^\alpha d\phi^\beta = {1\over  \alpha_h^2} \left[
{1\over  \varphi^2}\left(d\varphi^2 + |d{\vec\psi} |^2\right)\right]. 
\ee
We will also specialize to the case of non-zero but constant $V$, in 
which case we may set
\be\label{constantV}
V= \pm 2\alpha_h^2 m^2
\ee
for some constant $m^2$. In this case
\be
\Omega^2 = \alpha_h^2 m^2 \eta^2,
\ee
and hence
\be
\tilde G_{\mu\nu} d\Phi^\mu d\Phi^\nu
= m^2\left[- d\eta^2 + {\eta^2\over\varphi^2}\left( d\varphi^2 + |d\vec\psi|^2\right)\right] .
\ee
This is the Milne metric on Minkowski spacetime. In terms of the new coordinates 
$(u,v,{\vec x})$ defined by
\be
u= {\eta\over \varphi}, \qquad v= \eta\varphi + {\eta |{\vec \psi}|^2 \over\varphi}, 
\qquad \vec x = {\eta{\vec \psi} \over \varphi}, 
\ee
or,  equivalently, 
\be
\eta=\sqrt{uv-{\vec x}^2}\, , \qquad \varphi=u^{-1} \sqrt{uv-{\vec x}^2}\ , \qquad  
 {\vec \psi} ={ {\vec x}\over u}\, ,
\ee
one finds that
\be
\tilde G_{\mu\nu}\,  d\Phi^\mu d\Phi^\nu = m^2 
\left[-dudv + |d{\vec x}|^2\right] = 
m^2\, \eta_{\mu \nu}\,  dX^\mu dX^\nu,
\ee
where $X^\mu$ are standard cartesian coordinates ($2X^0= v+u$, $2X^1=v-u$) and $\eta_{\mu\nu}$
is the standard Minkowski metric in these coordinates. By a choice of origin for the 
affine parameter $\tau $,  any geodesic in this space can be written as
\be
X^\mu=  {1\over m^2}\, \eta^{\mu\nu} p_\nu \tau   + b^\mu, \qquad (p\cdot b=0)
\ee
for constant  mutually orthogonal $(N+1)$-vectors $p$ and $b$. The mass-shell (alias Friedmann) constraint is
\be
p^2 \pm m^2 =0,
\ee
which implies that we have a tachyon, and hence spacelike geodesics\footnote{It might appear strange that  for $V<0$ we would appear to be parametrizing a spacelike geodesic by a {\it time} parameter $\tau$. However. one must remember that 
$\tau$ has an interpretation as time in a different space to the space in 
which it parametrizes a geodesic.},  for $V<0$. Note that 
\be
\eta =  \sqrt{-X^2} = {1\over m}\sqrt{-m^2b^2 \pm \tau ^2}. 
\ee
The affine  parameter $\tau$  is related to the FLRW time $t$ by
\be
d\tau  =  |V|\eta\,  dt . 
\ee
To complete the possibilities for constant $V$ in this model, we may consider $V=0$. One again has a straight-line geodesic in the Minkowski coordinates:
\be
X^\mu =  p^\mu  \tau  + b^\mu, \qquad (p^2=0)
\ee
but we may no longer assume that $p\cdot b$ is zero. In fact it must be non-zero because if were zero then reality and positivity of $\eta$ would imply that  $b$ is  timelike, but a timelike vector cannot be orthogonal to a null vector. In addition, we now have
\be\label{zeroVtthat}
d\tau  = \eta \, dt . 
\ee

To determine the solution in each case as a function of the FLRW time $t$, we have only to integrate the relation for $d\tau$ in terms of $dt$. We consider in turn the three  cases: $V>0$,  $V<0$ and $V=0$. 

\begin{itemize}

\item $V>0$. In this case $p$ is timelike, so $b^2\ge0$ (with equality iff $b^\mu=0$) and
\be
{d\tau  \over \sqrt{\tau ^2 -m^2b^2}} = {V\over m}\, dt. 
\ee
Integration  yields
\be\label{integ}
\tau  = \cases{m\, e^{Vt/m} & $b=0$ \cr m\, |b|\cosh\left(Vt/m\right) & $b\ne0$,}
\ee
and hence
\be
\eta= \cases{ e^{Vt/m} & $b=0$ \cr |b|\sinh\left(Vt/m\right) & $b\ne0$.}
\ee
For $b=0$ we have the de-Sitter universe (as a flat FLRW universe). For $b\ne0$ we have a universe that begins with a big bang singularity at $t=0$ and then approaches a de Sitter universe at late times.
The singularity corresponds to the point at which the geodesic crosses
the Milne horizon, as shown in Fig 1.

\begin{figure}[h!]
\centering
\includegraphics*[width=200pt, height=180pt]{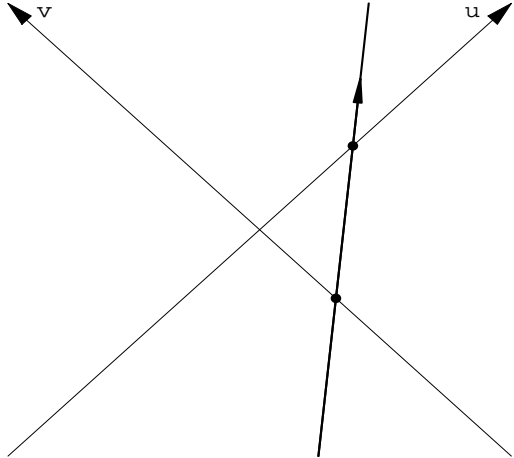}%
\caption{Geodesic for $V>0$. It describes an
expanding universe in the region $u>0,v>0$ 
with the singularity being at the point
 where the geodesic intersects the horizon $v=0$. }

\end{figure}

To complete the solution, we give the $t$-dependence  of the scalar fields. For $b=0$ they are constant, and for non-zero $b$ they are
\be
 \varphi = {\sinh(Vt/m)\over  {p^u\over m}
   \cosh(Vt/m)+ {b^u\over |b|}} \, , \qquad
   \vec \psi = { {\vec p\over m}
   \cosh(Vt/m)+ {\vec b\over |b|} \over  {p^u\over m}
   \cosh(Vt/m)+ {b^u\over |b|}} .
\ee
Note that although the full solution depends on many parameters, the spacetime geometry depends only on $|b|$. In particular, note the asymptotic behaviour
\be\label{etatau}
m\eta \sim \tau, 
\ee
which is valid for any $b$.

\item $V<0$. In this case $p$ is spacelike, so $b^2$ could be positive, negative or zero, but we must assume that $b^2<0$ in order to find a real solution for the scale factor. In this case
\be
{d\tau  \over \sqrt{m^2|b^2| - \tau ^2}} = {V\over m}\, dt.
\ee
Integration  yields
\be
\tau  = m|b| \cos(|V|t/m), 
\ee
and hence
\be
\eta = |b| \sin (|V|t/m). 
\ee
This represents a universe that begins with a big bang at $t=0$ and recollapses to a big crunch at $
t= m\pi/|V|$. The two singularities correspond to the two points at
which a spacelike geodesic in Minkowski space crosses the Milne
universe horizon, as shown in Fig.2. 

\begin{figure}[h!]
\centering
\includegraphics*[width=200pt, height=180pt]{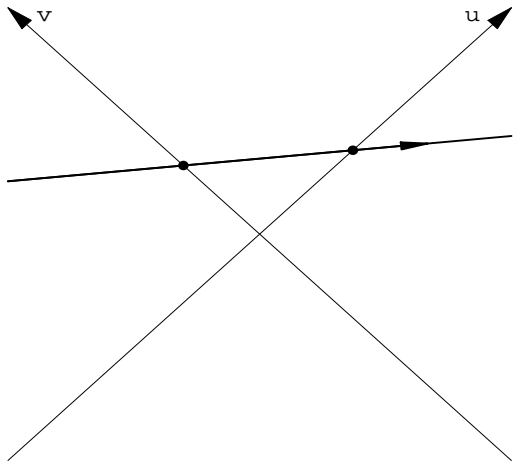}%
\caption{Geodesic for $V<0$. It describes a
 universe that begins with a big bang at the intersection point 
with the horizon $u=0$ and recollapses to a big
 crunch at the intersection point at $v=0$.}
\end{figure}

To complete the solution, we again give the $t$-dependence  of the scalar fields. 
\be
\varphi={ \sin(|V|t/m)\over  {p^u\over m}
   \cos(|V|t/m)+ {b^u\over |b|}}\, , \qquad 
   \vec \psi= { {\vec p\over m}
   \cos(|V|t/m)+ {\vec b\over |b|} \over  {p^u\over m}
   \cos(|V|t/m)+ {b^u\over |b|}}\, .
\ee

\item $V=0$. In this case
\be
{d\tau \over \sqrt{-b^2 - 2 (p\cdot b)\,  \tau }} = dt
\ee
and integration yields
\be
\tau  = -{b^2\over 2( p\cdot b)} - {(p\cdot b )\over 2}\,  t ^2. 
\ee
This  implies that
\be
\eta = |p\cdot b|\, t. 
\ee
The scalar fields are
\be
\varphi =  {\eta\over p^u\tau + b^u}, \qquad \vec\psi = {\vec p \tau + \vec b\over p^u\tau + b^u}
\ee
with $\eta$ and $\tau$ as  given. 

\end{itemize}

\noindent
Note how, in each case, the geometry of the Milne universe determines the cosmological singularies. With the exception of the $b=0$, $V<0$ case, there is always a big bang singularity at which $\eta \sim t$ and hence
\be\label{matdom}
S\sim t^{1\over d-1}.
\ee
This corresponds to the equation of state  of stiff matter ($P=\rho$). In other words, the FLRW universe 
is kinetic energy dominated near its singularities (and at all times if $V=0$). More generally, the formulae
\be
\alpha_h^2\, P = \left({\dot\eta\over \eta}\right)^2, \qquad 
\alpha_h^2\left(P+ \rho \right) = - \left({\eta\ddot\eta- \dot\eta^2\over \eta^2}\right)
\ee
can be used to determine $P$ and $\rho$ as functions of $t$. For example, one finds for constant positive $V$ that
\be
P=2 b^2V\left( {1-\sinh^2(Vt/m) \over \sinh^2(Vt/m)} \right)  \, ,\qquad
\rho=2 b^2V\left( {1+\sinh^2(Vt/m)\over \sinh^2(Vt/m)} \right). 
\ee

\section{A quantum Big-Crunch/Big-Bang transition} 

The cosmological models just described suggest a novel mechanism for the resolution of cosmological singularities because the geodesic evolution through these singularities in mini-superspace is smooth  in Minkowski coordinates; it is just that the scale factor becomes complex in the `non-Milne' regions of this space. We now elaborate this point with a particular case, and consider the implications of quantum mechanics. 

Consider a flat universe in the model of section {\ref{sect.hyper}
corresponding to a timelike geodesic in the  Minkowski moduli space
(the augmented target space) with $b\ne0$, as shown in Fig.3. 
\begin{figure}[h!]
\centering
\includegraphics*[width=200pt, height=180pt]{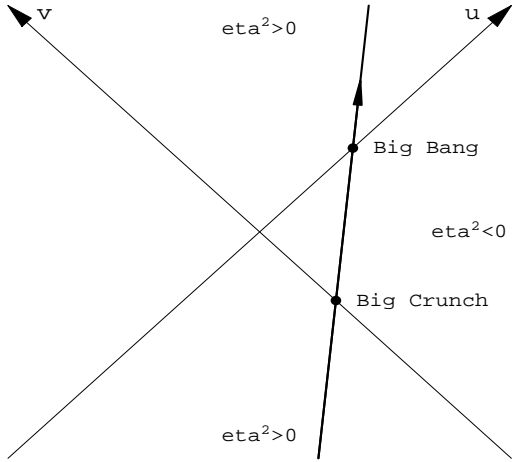}%
\caption{Transition from a big crunch to a big bang through a
`hidden' region where the scale factor is complex.}
\end{figure}
There are two Milne regions in this space, the past-Milne region ($u<0,v<0$) and the future-Milne region ($u>0,v>0$). A single timelike geodesic describes both a contracting universe in the past-Milne region and an expanding universe in the future-Milne region, corresponding to the following values of the affine time parameter:
\bea
{\rm past-Milne}: && \tau  < - m|b| \nn
{\rm future-Milne}: && \tau  >  m|b|. 
\eea
We earlier gave the solution only for the expanding universe in the future-Milne region, but the 
single trajectory in Minkowski space yields a solution in both Milne
regions\footnote{Note that  there is an ambiguity in the
 relation between the Minkowski space solution in terms of 
the cartesian coordinates $X$ and the Milne variables 
$\varphi, \vec\psi$, corresponding to $O(1,N)$ transformations
(which include $\varphi \rightarrow \varphi^{-1}$ when $\vec\psi=0$),
reflecting the freedom implied by isometries of the target space.}. Note that the section of the geodesic with
\be\label{hidseg}
|\tau | < m|b|
\ee
is not part of either the contracting universe in the past-Milne region nor the expanding universe in the future-Milne universe. It corresponds to a `hidden'  region of the Minkowski space (behind the Milne horizon) in which the scale factor is complex. In terms of the original gravity/scalar action  (\ref{gravscalar}), for which the scale factor is real, there is no way of connecting the past contracting phase with the future expanding phase--they are just different universes. However, our construction strongly suggests that they should be considered as part of a single universe that undergoes collapse to a singularity  followed by a re-expansion. Because this occurs in a `hidden'  region of the full moduli space of universes, it circumvents the theorem (see e.g. \cite{Khoury:2001bz}) that this cannot happen without violation of the null energy condition. 

We should stress that the issue under discussion here is {\bf not} how to resolve the cosmological singularity of the Milne universe, which is a mere coordinate singularity\footnote{In particular, we see no obvious connection to the work of \cite{Tolley:2002cv,Tolley:2003nx} on matching across the {\it real} singularity of an orbifold of Milne spacetime, although the methods used there may be relevant to our problem.}. The FLRW universes that we are considering are not Milne universes, and their cosmological singularities are not  coordinate singularities. The Milne `universe' appears as the moduli space of FLRW universes, and a given universe is a trajectory in this space; it just so happens that the cosmological singularities of these FLRW universes correspond to the points at which the trajectory crosses the Milne horizon.
In the past-Milne  region we have a collapsing universe approaching a big-crunch singularity while in the future-Milne region we have a universe expanding from a big-bang singularity. Near these singularities, and in between them, we should expect quantum mechanics to be relevant. 

To study the quantum mechanics of this big-crunch/big-bang transition we first restrict to the choice $N=2$; i.e.,  two scalar fields $(\varphi,\psi)$ which could  be considered as the dilaton and axion of some supergravity theory\footnote{The target space is therefore the coset space $Sl(2;\bR)/SO(2)$. 
String theory considerations motivate consideration of the compact hyperbolic space 
$Sl(2;\bZ)\backslash Sl(2;\bR)/SO(2)$, and the cosmological implications of this were considered in \cite{Russo:2003ky}.}. We quantize by imposing the phase-space constraint of the particle action (\ref{effL}) as an operator constraint on a Hilbert space; this gives us the Wheeler-DeWitt equation for 
a wave-function of the universe. With the natural operator ordering that yields the Laplacian on the Milne universe, this equation is
\be
\left[\Delta_H - c^2\eta^2 \right] \Psi = {\partial\over \partial\eta} \left(\eta^2 {\partial \Psi \over\partial\eta}\right), 
\ee
for wave-function $\Psi\left(\eta,\varphi,\psi\right)$, where
\be
\Delta_H = \varphi^2 \left({\partial^2 \over \partial\varphi^2} + {\partial^2 \over \partial \psi^2}\right)
\ee
is the Laplacian on the  two-dimensional hyperbolic space $H_2$ of unit radius, and we have set
\be
c^2 = {2V\over\alpha_h^2}.
\ee
For constant $V$, and hence constant $c$, the Wheeler-DeWitt equation can be solved by separation of variables:
\be\label{factorized}
\Psi = \Phi_\lambda\left(\varphi, \psi\right)\Xi_\lambda(\eta), 
\ee
where $\Phi_\lambda$ is an eigenfunction of the Laplacian on $H_2$, 
\be\label{htwoeigen}
\Delta_H \Phi_\lambda = \left(\lambda^2 -{1\over4}\right)\Phi_\lambda, 
\ee
and $\Xi$ satisfies the equation
\be\label{besseleq}
\left[\left(\lambda^2-{1\over4}\right) - c^2\eta^2 \right]\Xi = {\partial \over \partial \eta} \left( \eta^2 {\partial \Xi\over\partial\eta}\right). 
\ee
The general solution is
\be\label{besselsoln}
\Xi(\eta) = {1\over\sqrt{\eta}}\left[c_1 J_\lambda\left(c\eta\right) + c_2 Y_\lambda\left(c\eta\right) \right], 
\ee
for arbitrary constants $c_1,c_2$, where $J_\lambda$ is a Bessel function and
\be
Y_\lambda = \left[J_\lambda\cos\left(\lambda \pi \right) -J_{-\lambda}\right]/\sin\left(\lambda \pi\right).
\ee

Let us consider a wave-function of the following form
\be
\Psi = \sum_i {c_i\over\sqrt{\eta}}\left[ J_{\lambda_i}\left(c\eta\right) + i Y_{\lambda_i}\left(c\eta\right) \right]
\Phi_{\lambda_i} (\varphi,\psi)
\ee
for coefficients $c_i$ (we assume some standard normalization for the eigenfunctions of $\Delta_H$). 
From the asymptotic formula
\be
e^{i{\pi\over2}\left(\lambda + {1\over2}\right)} \sqrt{\pi\, z\over 2} \, \left[ J_\lambda\left(z\right) + i Y_\lambda \left(z\right)\right] \sim e^{iz}\left(1+ {\cal O}\left(1/z\right)\right],
\ee
we deduce that, for large $\lambda$, the wave-function takes the asymptotic form
\be\label{asym}
\Psi \sim   {e^{ic\eta}\over \eta} \, f(\varphi,\psi) 
\ee
for some function $f$ on $H_2$. The factor of $\eta^{-1}$ is expected because the measure in the integral  of $|\Psi|^2$ contains a factor of $\eta^2$. 
Given the validity of the semi-classical approximation  for large $\eta$ (which may depend on the nature of the function $f(\varphi,\psi)$ and hence on the choice of constants $c_i$) we expect 
\be
\Psi \sim e^{-iS}
\ee
for a slowly varying function $S$, which we can identify as the action of the classical trajectory. We see from (\ref{asym}), neglecting a logarithmic correction, that the $\eta$-dependence of this action is given by
\be
S \sim c\eta = 2m\eta, 
\ee
where we have used $V=2\alpha_h^2 m^2$. We saw earlier that $\eta \sim \tau/m$ for large $\eta$, so 
\be\label{HJ}
{V\over \alpha_h^2} {d\eta\over d\tau} = {\partial S \over \partial\eta} \, , 
\ee
as expected from Hamilton-Jacobi theory. 

The wave function can now be extended to a function of a complex variable $\eta$. As the Bessel functions are regular at the origin, the wave function has at most a singularity with a branch cut at $\eta=0$. One can take the cut (if one is needed) along the negative $\eta$ axis, so that there is no 
obstruction to an analytic continutation from real $\eta$ to pure imaginary $\eta$, passing around any singularity at $\eta=0$ in the complex $\eta$-plane. For large $|\eta|$ in the region where $\eta$ is pure imaginary, the wave-function is asymptotic to a sum of decaying and rising real exponentials. 
This suggests that the `hidden' region should be viewed as a classically forbidden region that a collapsing universe can pass through by quantum mechanical tunnelling into a region 
where it becomes an expanding universe.

Similar considerations apply for the case of $N$ scalar fields, in
which case
\be
\Xi (\eta ) = \eta ^{1-N\over 2}\left[ c_1 J_\lambda (c\eta )+ c_2
Y_\lambda(c\eta ) \right], \qquad 
\Delta_H \Phi= \left[\lambda^2 - {1\over4}\left(N-1\right)^2\right] \Phi.
\ee 
This leads to a wavefunction $\Psi$ with similar asymptotic behaviour
for all $N$.

\section{Non-flat cosmologies}

For $k\ne0$ the geodesic 
interpretation  of \cite{Townsend:2004zp} required consideration of 
a `doubly-augmented' target space with coordinates $\Psi^A=(\Phi^\mu, \psi_*)$ and metric
\be
G_{AB}\, d\Psi^A d\Psi^B = G_{\mu\nu}d\Phi^\mu d\Phi^\nu - k\, 
C^{-1}  \eta^{-2(d-2)/(d-1)}\,  d\psi_*^2.
\ee
where $C$ is any non-zero constant. Let $\Pi_A=(p_\mu, p_*)$ be the variables canonically conjugate to $\Psi^A$, and consider the particle mechanics Lagrangian 
\be\label{nonzerolag}
L= \Pi_A \dot\Psi^A -{1\over2}e \left(\eta^{-2} G^{AB}\Pi_A\Pi_B + 2V\right) - \chi\left(p_* +  k\right).
\ee
The new constraint imposed by $\chi$ is associated with a new gauge
invariance, for which the non-zero transformations
are
\be
\delta\psi_* = f, \qquad \delta \chi = \dot f
\ee
for
arbitrary function $f$. This allows us to choose a gauge for which
$\chi=0$. If we also fix the time-reparametrization invariance by the
gauge choice (\ref{gaugechoice}) then we can 
eliminate $\Pi_A$ by its equation of motion
\be
\Pi_A = 2 \tilde G_{AB}\dot \Psi^A, \qquad 
\tilde G_{AB} := \Omega^2 G_{AB}, 
\ee
where  $\Omega^2$ is the same conformal factor as before.  The Lagrangian becomes
\be
L= \tilde G_{AB}\dot \Psi^A\dot\Psi^B, 
\ee
so the trajectories are affinely parametrized geodesics. The mass-shell constraint is
\be
G_{AB}\dot \Psi^A\dot\Psi^B  = \cases{-{\rm sign}\, V & $V\ne 0$ \cr 0 &$V=0$}
\ee
and the $\chi$ constraint  is equivalent to
\be
\dot\psi_* = {C\over 2\Omega^2} \, \eta^{2(d-2)/(d-1)}. 
\ee
For the choice $C= (d-1)(d-2)/2$, these equations are precisely the ones shown in \cite{Townsend:2004zp} to describe general cosmological trajectories (this is also true for any other choice of $C$). Thus, we have found a particle mechanics reformulation, via the Lagrangian (\ref{nonzerolag}), of the construction of \cite{Townsend:2004zp}. When $k=0$, solving the constraint imposed by $\chi$ returns us to the Lagrangian (\ref{effL}) already shown to be the effective particle Lagrangian in this case. When $k\ne0$, the constraint imposed by $\chi$ implies a particle mechanics version of Scherk-Schwarz-type dimensional reduction (a discussion of which can be found in \cite{deAzcarraga:1992de}), thereby providing a simple geometric interpretation of the projection needed in this case. 

An alternative formulation of $k\ne0$ trajectories as geodesics can be deduced  from
(\ref{nonzerolag}) by solving the $\chi$ constraint,  and discarding a total derivative,  to arrive at the alternative effective Lagrangian
\be\label{effL2}
L= p_\mu \dot \Phi^\mu - {1\over2}e\left(\eta^{-2}G^{\mu\nu} p_\mu p_\nu + 2\, V_{eff}\right), 
\ee
where the `effective potential'  is
\be
V_{eff} =V - {1\over2}k(d-1) \eta^{-2/(d-1)}.
\ee
The previous discussion for $k=0$ cosmologies can now be carried over to the $k\ne0$ case but in terms of the effective potential, which now depends on the scale factor as well as the scalar fields. 
Assuming that $V_{eff}$ is non-zero, one has 
\be
\tilde G_{\mu\nu} = {1\over2}\eta^2|V_{eff}| G_{\mu\nu}\, ,\qquad d\tau = \eta|V_{eff} |dt. 
\ee
The gauge-fixed effective Lagrangian is therefore
\be
L= {|V_{eff}|\over 2\alpha_h^2}\left[ -\dot\eta^2 + \alpha_h^2 \eta^2 G_{\alpha\beta}\dot\phi^\alpha\dot\phi^\beta\right], 
\ee
where the overdot indicates differentiation with respect to $\tau$. The constraint is
\be\label{altcon}
{1\over2}V_{eff} \left(-\dot\eta^2 + \alpha_h^2 \eta^2 
G_{\alpha\beta}\dot\phi^\alpha\dot\phi^\beta\right) =-1. 
\ee
In the special case that  $\dot\phi^\alpha=0$, this constraint is equivalent to 
\be
\left(d-1\right) \, {dS\over dt} = \sqrt{2VS^2 -k\left(d-1\right)}. 
\ee
Clearly,  the assumption $\dot\phi^\alpha=0$ is consistent only if
the right hand side is real. One case where it {\it is} consistent is
if $k=-1$ and $2V=-(d-1)^2\omega^2$ for  constant $\omega$. One then finds that
\be
S(t) = {1\over \sqrt{(d-1)\omega^2}} \, \sin\omega t , 
\ee
which yields anti-de Sitter space, as one would expect. 

As a further special case, we now consider models  with $V=0$, for which
\be
V_{eff} = - {1\over2}k(d-1) \eta^{-2/(d-1)}. 
\ee
The geodesic trajectories are therefore null for $k=0$, timelike for $k=-1$ and spacelike for $k=1$.  The affine parameter $\tau$ is such that
\be\label{tthat2}
d\tau = {1\over2}\left(d-1\right) \eta^{(d-3)/(d-1)} dt . 
\ee
If we now choose a hyperbolic target space with metric\footnote{Note that the radius differs from our previous choice, and we take $N=2$ for simplicity.}
\be
G_{\alpha\beta}\, d\phi^\alpha d\phi^\beta = {\alpha_h^2 \over 4 \varphi^2} \left(d\varphi^2 + |d\psi|^2\right), 
\ee
and define 
\be
\zeta = \eta^{(d-2)/(d-1)} = S^{d-2},
\ee
then we find that the gauge-fixed Lagrangian is 
\be
L \propto \left[-\dot\zeta^2 + {\zeta^2\over\varphi^2}\left( \dot\varphi^2 + \dot\psi^2 \right)\right].
\ee
The particle trajectories are again geodesics in the Milne metric on a region of Minkowski space and we may introduce cartesian coordinates $X^\mu$ as before, in terms of which the geodesic is 
\be
X^\mu = {4\over \alpha_h^2}\, \eta^{\mu\nu}p_\nu \tau  + b^\mu, 
\ee
with  the mass-shell constraint 
\be
p^2 = {1\over4} k(d-1)\alpha_h^2. 
\ee
If we assume that $k\ne0$ (because otherwise $V_{eff}=0$ and we have 
a case considered earlier) then we may also assume that $p\cdot b=0$,
in which case
\be
\zeta^2 \equiv -X^2 = -{4(d-1)k\over \alpha_h^2}\, \tau ^2 -b^2. 
\ee
From (\ref{tthat2}) we have
\be
d\tau  = {1\over2}\left(d-1\right) \zeta^{(d-3)/(d-2)} dt\ .
\ee
The determination of the function $\tau (t)$ is no longer quite so
simple, but this is just a matter of the {\it parametrization} of 
the cosmological trajectory. The Milne geometry itself provides us 
with the essential qualitative features. For $k=1$ we have a spacelike 
geodesic which (if it yields a solution with real scale factor) must 
intersect the Milne horizon twice, thus yielding a universe that 
begins with a big bang and ends with a big crunch (although the 
trajectory in Minkowski spacetime containing the Milne spacetime 
can be smoothly continued through these cosmological singularities). 
For $k \le0$ we have a timelike or null geodesic which can intersect 
the future Milne horizon once only; this intersection is the big 
bang singularity of a universe that expands forever.

\section{Comments}

In this paper we have explained how some results of \cite{Townsend:2004zp}, in which cosmological solutions of gravity/scalar  models with arbitrary scalar potential were shown to have an interpretation as geodesics in an appropriate metric on the augmented target space, follow simply from a reformulation of cosmology as relativistic particle mechanics. The construction of \cite{Townsend:2004zp} was straightforward for flat ($k=0$) cosmologies but less so for non-flat ($k\ne0$) cosmologies involving, as it did, a projection from a geodesic in a `doubly-augmented'  target space. The reformulation  presented here considerably clarifies the geometrical meaning of the latter construction. The doubly-augmented target space is a $U(1)$ bundle over the augmented target space, and the projection to the latter is just a Scherk-Schwarz dimensional reduction on the $U(1)$ fibre, at least for $k=-1$ for which the extra dimension is spacelike. For $k=1$ the extra dimension is timelike and the physical interpretation is less appealing, suggesting that $k=1$ universes are somehow unphysical\footnote{Note that the Schwarzschild radius of matter within the cosmological (particle) horizon of an observer in  a decelerating $k=1$ universe is {\it less} than  the horizon radius!  Thus, in a sense, such an observer is inside a black hole, although it is not really correct to identify the surface at the Schwarzschild radius as an event horizon in this context. }, although it  is perhaps appropriate to recall here  that  two-time particle mechanics models have found other applications \cite{Bars:2000qm}.  In fact, as we have shown here, it is possible to consider $k\ne0$ cosmologies as geodesics in the augmented target space by attributing the effects of spatial curvature to a contribution to an effective potential. This is actually a special case of an even more general construction (see e.g. \cite{Greensite:1995ep}) based on the Maupertuis-Jacobi principle of classical mechanics. We used this analogy to argue that cosmological singularities are boundaries  of  classically forbidden regions in the space of all FLRW universes. 

We have applied these ideas to various models with hyperbolic scalar target spaces and constant potentials. A feature of the chosen models is that  the metric on the augmented target space (mini-superspace) with respect to which cosmological solutions (of specified curvature) are geodesics is the Milne metric. In other words, the Milne `universe' is the moduli space of FLRW universes in these models. This allows all solutions (or, at least, trajectories) to be found exactly. This state of affairs arises for $k=0$ cosmologies with a constant potential (as also noted by Waldron  in the context of $(2+1)$-dimensional gravity \cite{Waldron:2004gg}) and for $k=\pm 1$ cosmologies with zero potential. We have argued, in the context of $k=0$ cosmologies with a positive cosmological constant, that  there is a natural quantum big-crunch/big-bang transition that corresponds to a geodesic trajectory that passes through a region in moduli space that is behind  the Milne horizon. We have argued that the quantum analogue of the absence of a classical obstruction in moduli space is analyticity of the wave-function as a complex function of the variable $\eta$ (which is itself a power of the scale factor). This variable is real in the past and future Milne regions and imaginary elsewhere. Analyticity of the wave function allows a continuation into the `hidden' region of Minkowski spacetime, which is analogous to a classically forbidden region in a problem of quantum mechanical tunnelling through a barrier. 

We should also stress that the existence of such a horizon in moduli space is itself a very special 
model-dependent feature. It depends, obviously, on the existence of scalar fields, and on the choice of a hyperbolic target space for these fields. Less obviously, it also depends on the precise choice of radius of this hyperbolic space; a different choice  would not yield a Milne universe moduli space (at least not one with respect to which the cosmological trajectories are geodesics), and  a conical singularity would replace the Milne horizon. Effectively, this means that the distance scale determined by the dimensionful sigma-model coupling must be a definite numerical factor times the Planck scale, so the scalar fields should be considered as `gravi-scalars'. This occurs naturally in many supergravity theories, as do hyperbolic target spaces, which arise as moduli spaces in toroidal compactification. Whether there is any supergravity theory that precisely realizes our construction is an interesting open question.

\bigskip
\noindent
{\bf Acknowledgements.} We thank Eric Bergshoeff and Mattias Wohlfarth  for  discussions on the 
particle Lagrangian formulation of cosmology, and Lee Smolin for helpful correspondance. J.R. acknowledges  partial support by
the European Commission RTN programme under 
contract HPNR-CT-2000-00131 and
by MCYT FPA 2001-3598 and CIRIT GC 2001SGR-00065.

\end{document}